\documentstyle[epsfig,aps]{revtex}

\def\be{\begin{eqnarray}}
\def\ee{\end{eqnarray}}
\def\nn{\nonumber\\ }
\def\Tr{{\rm Tr}\;}
\def\im{{\em Im}\;}
\def\re{{\em Re}\;}

\def\MeV{{\rm \; MeV}}

\def\T{{\rm T}}
\def\hS{\hat {\cal S}}

\def\F{{\rm\bf F}}

\def\bH{{\rm\bf H}}
\def\J{{\rm\bf J}}
\def\R{{\rm\bf R}}
\def\V{{\rm\bf V}}
\def\W{{\rm\bf W}}
\def\j{{\rm\bf j}}

\begin{document}


\title{Dilepton and Photon Emission Rates from a Hadronic Gas III}

\author{C.-H. Lee$^a$, H. Yamagishi$^b$ and I. Zahed$^a$}
\address{a) Department of Physics \& Astronomy, SUNY at Stony Brook, 
Stony Brook, NY 11794, USA\\
b) 4 Chome 11-16-502, Shimomeguro, Meguro, Tokyo, 153, Japan}  

\maketitle

\begin{abstract}
We extend our early analyses of the dilepton and photon emission rates 
from a hadronic gas to account for strange mesons using  a density
expansion. The emission rates are reduced to vacuum correlation functions
using three-flavor chiral reduction formulas, and the latters are assessed
in terms of empirical data. Using a fire-ball, we compare our results to the
low and intermediate mass dilepton data available from CERN. Our results
suggest that a baryon free hadronic gas does not account for the excess of
low mass dielectrons observed at CERES 
but do well in accounting for the intermediate
dimuons at HELIOS. The same observations apply to the recent low and high 
$p_t$ dielectron rates from CERES.

\end{abstract}

\section{Introduction}
\label{sec:1}

The latest experiments at the CERN SPS~\cite{CERES} machine have revealed a 
sizable enhancement of low mass dileptons above the two-pion threshold, 
triggering a number of theoretical 
investigations~\cite{ALL,JIM,JIM2,LKB96,RAPP}. A
somewhat smaller enhancement was also noticed in the intermediate mass
region around the phi both in the dielectron (CERES) and dimuon (HELIOS) 
experiments~\cite{CERES,ABOVE}. This region requires a strangeness assessment
of the emission rates. An example is the recent analysis by Li and 
Gale~\cite{GALE}. 

In a recent series of investigations we have assessed the dilepton and 
emission rates emanating from a hadronic gas without strangeness using 
general principles~\cite{JIM,JIM2}. 
We have found that the current low mass dilepton
enhancement can only be accounted for by allowing for a sizable nucleon
density~\cite{JIM2}. 
It is important to stress that our construction is not a model.
It relies on the strictures of broken chiral symmetry, unitarity and data.
Most model calculations should agree with our analysis to leading order 
in the pion and nucleon densities. An example being the comparison
by Gale for the baryon free rates~\cite{CHARLES}.

The aim of the present work is to extend our baryon free analysis to the
strangeness sector, to account for strange mesons in the hadronic gas. Our
results will borrow on the extension of the chiral reduction formulas to
QCD with three flavors~\cite{LEEZA}. To leading order in the
meson densities, the emission rates in the hadronic gas involve forward
scattering amplitudes of real (photon) and virtual (dilepton) photons, 
which behaviour is constrained by data mostly from electro-production  and 
photon-fusion reactions. By including strangeness we aim at evaluating the
emission rates in the intermediate mass region around the phi. Throughout
we will not consider baryons.

The structure of the paper is as follows : in section~\ref{sec:2}, 
we derive the emission
rates for a baryon free hadronic gas to leading order in the final meson 
densities including strange mesons. In section~\ref{sec:3}, 
we discuss the relevance of
the various contributions and introduce the pertinent spectral functions.
In section~\ref{sec:4}, we discuss the integrated dielectron and
photon emission rates. 
In section~\ref{sec:5}, 
we use a fire ball scenario to account for our time-evolved rates in 
comparison to current CERES and HELIOS data. 
In section~\ref{sec:6}, we discuss the current $p_t$ spectrum of CERES data. 
Our conclusions are summarized in 
section~\ref{sec:7}.

\section{Dilepton Rates}
\label{sec:2}

In a hadronic gas in thermal equilibrium, the rate ${\bf R}$ of dileptons 
produced in an unit four volume is directly related to the electromagnetic
current-current correlation function~\cite{LARRY,WELDON}. For massive 
dileptons $m_{1,2}$ with momenta $p_{1,2}$, the rate per unit invariant 
momentum $q=p_1+p_2$ is 
     \be
     \frac{d \R}{d^4 q} &=&-\frac{\alpha^2}{6\pi^3 q^2}
     \left(1+\frac{2 m_l^2}{q^2}\right)
     \left(1-\frac{4 m_l^2}{q^2}\right)^{1/2} \W (q)
     \ee
where $\alpha=e^2/4\pi$ is the fine structure constant, and
     \be
     \W (q) &=& \int d^4 x e^{-iq\cdot x} \Tr 
        \left( e^{-(\bH-\F)/T} \J^\mu(x) \J_\mu(0)\right)
        = \frac{2}{1+e^{q^0/T}}\im \W^F (q) \ .
     \label{eq:2}
     \ee
Here $e\J_\mu$ is the hadronic part of the electromagnetic current,
${\bf H}$ is the hadronic Hamiltonian, $\F$ is free energy, $T$ is the
temperature, and $\W^F(q)$ is
     \be
     \W^F(q) &=& i\int d^4 x e^{iq\cdot x}\Tr
      \left( e^{-(\bH-\F)/T} \T^\star \left(\J^\mu(x) \J_\mu(0)\right)\right)
     \nn
      &=& i\int d^4 x e^{iq\cdot x}
      \langle 0 |\T^\star( \J^\mu(x)\J_\mu(0)|0\rangle
     \nn
      && +\sum_a i\int \frac{n^a(\omega^a_k)}{2\omega^a_k}
      \int d^4x e^{iq\cdot x}
      \langle\pi_{in}^a(k) |\T^\star (\J^\mu(x)\J_\mu(0)) 
                      |\pi^a_{in}(k)\rangle_{con.}
     \nn
      && +\cdots\label{expansion}
     \ee
where the sum is over physical mesons including strange ones. For temperatures
$T\leq m_{\pi}$ the first two terms in Eq.~(\ref{expansion}) are dominant. 
The first term relates to the vacuum current-current correlator and 
captures the essentials of the resonance gas model. It follows from $e^{+}e^-$ 
annihilation data. The second term is the first correction to the resonance
gas model resulting from one meson in the final state. Two and higher meson
corrections in the final state can be evaluated using similar 
arguments~\cite{JIM}. 

Using the definition of the electromagnetic current
$\J_\mu^{em} = \bar q\gamma_\mu Q q = \V_\mu^3 +\frac{1}{\sqrt 3} \V_\mu^8$,
and the decomposition
     \be
     T^\star (\J_\mu {\J_\nu}) &=&
      T^\star (\V^3_\mu {\V^3_\nu} )
     + \frac{1}{3} T^\star (\V^8_\mu {\V^8_\nu} )
     + \frac{1}{\sqrt 3} T^\star (\V^3_\mu {\V^8_\nu}+\V_\nu^3\V_\mu^8 ),
     \ee       
we may follow~\cite{JIM,YAZA} and rewrite Eq.~(\ref{expansion}) in terms 
of two-point correlation functions
     \be
     \im \left( i\int_y e^{-iq\cdot y}
       \left\langle 0 \left| \hS \T^\star
       \left(\V_\mu^c(y)\V_\nu^d(0)\right)\right|0 \right\rangle \right)
      &=& \left(-q^2 g_{\mu\nu}+q_\mu q_\nu\right)\im \Pi_V^{cd} (q^2) \nn
     \im \left( i\int_y e^{-iq\cdot y}
       \left\langle 0 \left| \hS \T^\star
       \left(\j_{A \mu}^c(y)\j_{A \nu}^d(0)\right)\right|0 \right\rangle \right)
      &=& \left(-q^2 g_{\mu\nu}+q_\mu q_\nu\right)\im \Pi_A^{cd} (q^2),
     \label{spectral}
     \ee
and additional three- and four-point correlation functions using
three-flavor chiral reduction formulas~\cite{LEEZA}. The axial-vector current
$\j_A$ appearing in Eq.~(\ref{spectral}) is one pion reduced~\cite{JIM,YAZA}.

With the above in mind, we have
     \be
     \im\W^F(q) =-3q^2\im\left(\Pi_V^{I}(q^2)+\frac 14 \Pi_V^{Y}(q^2)\right)
      +\int \frac{d^3k}{(2\pi)^3} \W_1^F(q,k),
     \label{zero}
     \ee
with $\Pi_V^{I} \equiv \Pi_V^{33}$ and $\Pi_V^{Y} \equiv \frac 4 3\Pi_V^{88}$.
The first term is the analogue of the resonance gas contribution with no
chiral reduction involved. The second term is the correction to the resonance
gas model resulting from one meson in the final state. Use of the on-shell
three flavour chiral reduction formulas gives for the part involving solely
the two-point correlators
     \be
      \W^F_1(q,k) &=&
      g^\pi_k \frac{12}{f_\pi^2} q^2\im \Pi_V^{I}(q^2)
     + g^K_k \frac{12}{f_K^2}  q^2\im \left( \Pi_V^{I}(q^2) 
      + \frac 34 \Pi_V^{Y}(q^2) \right)
     \nn
     &&
      - g^\pi_k \frac{6}{f_\pi^2} (k-q)^2 \im\Pi_A^{I} ((k-q)^2)
      - g^K_k \frac{6}{f_K^2} (k-q)^2
      \left(\im\Pi_A^{V}((k-q)^2) +\im\Pi_A^{U}((k-q)^2) \right) \nn
     && 
      - g^\pi_k \frac{6}{f_\pi^2} (k+q)^2 \im\Pi_A^{I} ((k+q)^2)
      - g^K_k \frac{6}{f_K^2} (k+q)^2
      \left(\im\Pi_A^{V}((k+q)^2) +\im\Pi_A^{U}((k+q)^2) \right) 
     \nn
     && +
      g^\pi_k \frac{8}{f_\pi^2} (\nu^2-m_\pi^2 q^2)
        \re\left(\tilde \Delta_R^\pi (k+q)+\tilde \Delta_R^\pi (k-q) \right)
        \im \Pi_V^{I}(q^2) \nn
     &&
      + g^K_k \frac{8}{f_K^2} (\nu^2-m_K^2 q^2)
        \re\left(\tilde \Delta_R^K (k+q)+\tilde \Delta_R^K (k-q) \right)
        \im \left( \Pi_V^{I}(q^2) +\frac34\Pi_V^{Y}(q^2) \right), 
     \label{two}
     \ee
where $\nu=k\cdot q$. The contributions from three- and four-point 
correlators read
     \be
      \W^F_1(q,k) &=&
     \left\{
      g^\pi_k  3\sqrt 3 \frac{f_\eta m_\eta^2}{f_\pi^2}
              \frac{\hat m}{\hat m+2 m_s} 
       -g^K_k \sqrt 3 \frac{f_\eta m_\eta^2}{f_K^2}
              \frac{\hat m+m_s }{\hat m+2 m_s} 
       -g^\eta_k \frac{\sqrt 3}{3} \frac{m_\eta^2}{f_\eta}
     \right\} 
     \im \tilde G_\sigma^8
     \nn
     &&
     -\frac 23 \frac KC \left\{ g^\pi_k \frac{3 \hat m}{f_\pi^2}
     +g^K_k \frac{2 (\hat m+m_s)}{f_K^2}+g^\eta_k
            \frac{(\hat m+2m_s)}{3 f_\eta^2} \right\}
     \im  \tilde G_\sigma^0 \nn
     &&
     - g^a_k  k^\beta k^\alpha (E^{-2})^{aa} 
     \im \left\{ i
      \int_x \int_y \int_{z}
      e^{-ik\cdot z -iq\cdot y+i q\cdot x} \right. \nn
     && \left.
       \left\langle 0\left|\hS \T^\star \left[
           \left(\V^{\mu,3}(x)+\frac{1}{\sqrt 3}\V^{\mu,8}(x) \right)
           \left(\V_\mu^3(y)+\frac{1}{\sqrt 3}\V_\mu^8(y) \right)
           {\j_A}_\beta^a(z){\j_A}_\alpha^a(0)\right] \right| 0\right\rangle
      \right\} \nn
     &&
     - g^a_k k^\beta \left( f^{a3l} +\frac{1}{\sqrt 3} 
        f^{a8l}\right) (E^{-2})^{aa}
     \im \left\{ 
      \left[\delta_\alpha^\mu -(q+2k)^\mu (k+q)^\alpha\tilde\Delta_R^l (k+q)
      \right] \phantom{\frac 13} \right.  \nn 
     && \;\;\;  \left. \left( i \int_y \int_{z}
      e^{-ik\cdot z -i q\cdot y}
        \left\langle 0\left|\hS \T^\star \left[ 
           \left(\V_\mu^3(y) +\frac{1}{\sqrt 3} \V_\mu^8(y) \right)
           {\j_A}_\beta^a(z){\j_A}_\alpha^l(0) \right] \right| 0\right\rangle
      \right) \right\} \nn 
     && + {\rm permutations} (q\rightarrow -q, k\rightarrow -k) \ ,
     \label{three}
     \ee
where the permutation applies only for the last term,
and  no-mixing is assumed ( $\Pi_V^{38}=\Pi_V^{83}=0$ ). The thermal
meson density function $g^a_k$ and the meson propagator $\tilde \Delta_R^a (q)$
are defined by
     \be
     g^a_k \equiv \frac{n^a(\omega^a_k)}{2\omega^a_k}
           =\frac{1}{2\omega^a_k} \frac{1}{e^{\omega^a_k/T}-1}\ ,
     \;\;\;\;\;
     \tilde\Delta_R^a (q) \equiv \frac{1}{q^2-m_a^2 +i\epsilon}.
     \ee
The three point correlator $\tilde G_\sigma^h$ is given as
     \be
     \tilde G_\sigma^h &=&
     \int_z \int_y e^{iq\cdot (z- y)}
      \left\langle 0\left|\hS \T^\star \left[ 
         \left( \V^{\mu,3}(z) + \frac{1}{\sqrt 3}\V^{\mu,8}(z) \right)
         \left( \V_\mu^3(y) + \frac{1}{\sqrt 3}\V_\mu^8(y)  \right)
       \sigma^h(0)\right] \right| 0\right\rangle.
     \ee
The meson decay constants,
$f_K\approx 1.24 f_\pi$, $f_\eta \approx 1.32 f_\pi$,
$f_{\eta^\prime}\approx 0.74 f_\pi$ are used~\cite{shuryak93}.
The results Eqs.~(\ref{zero}-\ref{three}) reduce to the two-flavor results
discussed in~\cite{JIM}. Since they include strangeness they can be used
all the way through the phi region.

The spectral functions appearing in Eqs.~(\ref{zero},\ref{two}) are
related to $e^{+}e^{-}$ annihilation data. They will be borrowed from
experiment. The three- and four-point correlation functions 
$\V\V\j_A\j_A$, $\V\j_A\j_A$ and $\V\V\sigma$ are constrained
by the two-photon fusion reactions and crossing symmetry~\cite{LEEZA}.
A detailed analysis of these processes show that their contribution to
our rates can be ignored for both CERES and HELIOS (2\% correction).
These observations
confirm and extend the ones made in~\cite{JIM} to the three-flavour case.
Hence, the two- and three-point contributions will be omitted for most of the
discussion to follow. 

\section{Spectral Functions}
\label{sec:3}

In this section we derive explicit expressions for the spectral functions.
The empirical information on the two-point correlators will be inserted 
in the rate production by means of suitable spectral weights. For instance,
the contribution of the non-strange hadrons to the spectral weights is 
dominated by the $\rho$ ($m_\rho = 768.5 \MeV,\; \Gamma_{0,\rho} = 150.7 \MeV$)
and the $a_1$ ($m_{a_1} = 1230 \MeV,\; \Gamma_{0,a_1} = 400 \MeV $)
resonances. For that, we will follow the arguments of~\cite{JIM} and use
the following parametrizations
     \be
     \Pi^I_V (q^2) &=& \frac{ f_\rho^2}{q^2}
        \frac{m_\rho^2+\gamma q^2}{m_\rho^2-q^2-i m_\rho\Gamma_\rho (q^2)}
     \nn
     \Pi^I_A (q^2) &=& \frac{f_{a_1}^2}{m_{a_1}^2-q^2-im_{a_1} 
                       \Gamma_{a_1}(q^2)}
     \ee
with $f_\rho=\sqrt 2 f_\pi$ and $f_{a_1} = 190 $ MeV.
The decay widths $\Gamma$ are given by
     \be
     \Gamma_\rho (q^2) &=& \theta(q^2-4 m_\pi^2) \Gamma_{0,\rho}
          \frac{m_\rho}{\sqrt{q^2}}
         \left(\frac{q^2-4 m_\pi^2}{m_\rho^2-4 m_\pi^2}\right)^{3/2} 
     \nn
     \Gamma_{a_1} (q^2) &=& \theta(q^2-9 m_\pi^2) \Gamma_{0,a_1} 
         \frac{m_{a_1}}{\sqrt{q^2}}
         \left(\frac{q^2-9 m_\pi^2}{m_{a_1}^2-9 m_\pi^2}\right)^{3/2}
     \ee
where $\theta(x)$ is Heaviside functions (step function). 
Analogous parametrizations will be used for the strange 
resonances as well including the $\omega$ and $\phi$
in $\Pi_V$, and $K_1$ in $\Pi_A$. 
The isovector part of the spectral weight $\Pi_V^I$ is dominated by 
the $\rho$ and its radial excitations, while the hypercharge part 
$\Pi_V^Y$ is dominated by the omega, the phi and their radial excitations.
In the physical basis 
     \be
      \omega_8 &=& \sqrt{\frac 13}\omega-\sqrt{\frac 23}\phi \nn
      \Pi_V^{88} &=& \frac 13 \Pi_V^\omega +\frac 23 \Pi_V^\phi \ .
     \ee

\begin{figure}
\centerline{\epsfig{file=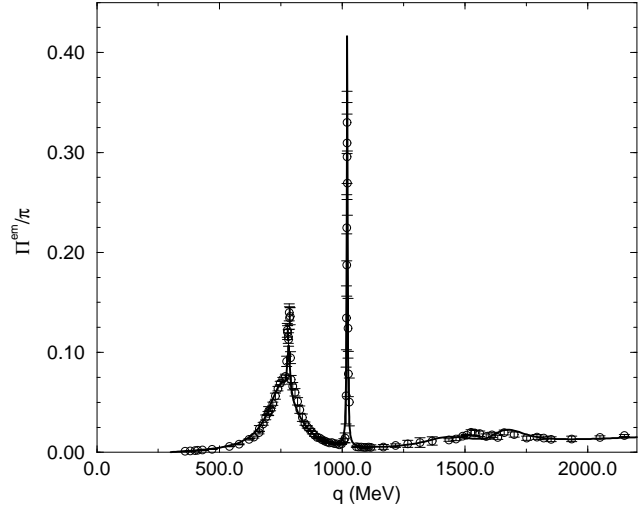,height=8cm}}
\caption{Electromagnetic Spectral Function. See text}
\label{fig-em}
\end{figure}

In Fig.~\ref{fig-em} we show the electromagnetic spectral function 
$\Pi_V^{em} \equiv \Pi_V^I+\frac 14 \Pi_V^Y$ following from our
parameters (solid curve) in comparison with the data compiled in~\cite{Huang}. 

The resonance parameters
we have used are summarized in Table.~\ref{tab:1}~\cite{databook}.
The decay constants are fit to the empirical spectral 
weights~\cite{shuryak93}. They are all within 10 \% of the constituent
quark model. 
Although the axial-vector spectral weight is not well-known at large 
invariant mass, we observe that its numerical contribution to 
the dilepton emission rates is overall small. 

\begin{table}
\begin{tabular}{clcccc}
  &  & $I^G(J^{PC})$ & mass ($m_i$) & decay width ($G_i$) 
  & decay constant ($f_i$) \\
\hline
\hline
 $\Pi_V^I$  & $\rho (770)$  & $1^+(1^{--})$ & 768.5 & 150.7 & 130.67  \\
            & $\rho (1450)$ &               & 1465  &  310 &  106.69 \\
            & $\rho (1700)$ &               & 1700  &  235 &  75.44 \\
\hline 
 $\Pi_V^Y$  & $\omega (782)$ & $0^-(1^{--})$ & 781.94 & 8.43 & 46 \\
            & $\omega (1420)$ &              & 1419 & 174& 46\\
            & $\omega (1600)$ &              & 1649 & 220& 46\\
\cline{2-6}
            & $\phi (1020)$ & $0^-(1^{--})$ & 1020 & 4.43  & 79 \\
            & $\phi (1680)$ &               & 1680 & 150  &  79\\
\hline
\hline
 $\Pi_A^I$ & $a_1 (1260)$ & $1^-(1^{++})$ & 1230 & 400 & 190 ($f_\rho$) \\
\hline
 $\Pi_A^{UV}$ & $K_1 (1270)$ & $\frac 12(1^+)$ & 1273 &  90 & 90 \\
              & $K_1 (1400)$ &                 & 1402 & 174 & 90 \\
\end{tabular}
\caption{Resonance parameters}
\label{tab:1}
\end{table}

\begin{figure}
\centerline{\epsfig{file=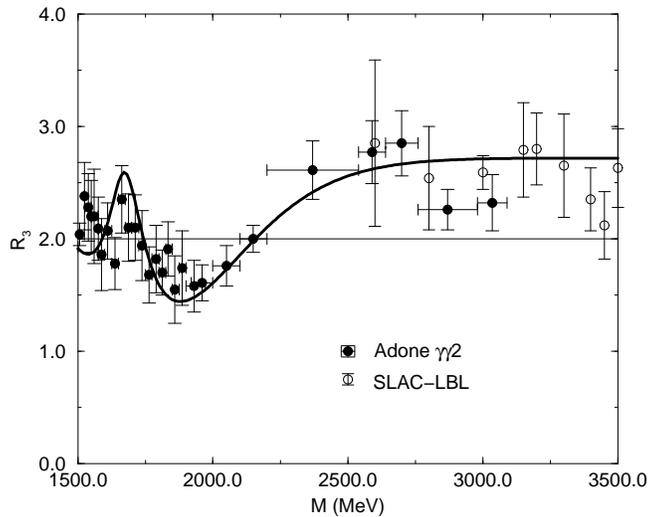,height=8cm}}
\caption{$R_{\ge 3}$ values. Data from Adone~\protect\cite{Bacci}
and SLAC-LBL~\protect\cite{Siegrist}.}
\label{fig:R3}
\end{figure}

The high energy tail of the vector spectral weight is still above the free
$q\overline{q}$
threshold well through the high mass region as shown in Fig.~\ref{fig:R3} for
$R_{\ge 3}=12\pi \,{\rm Im}\Pi (s)$ without the $J/\Psi$ contribution. We
will use the parametrization 
     \be
        \frac{0.9}{8\pi} 
       \left(1+\tanh\left(\frac{\sqrt{q^2}-q_1}{\tilde q}\right)\right)
     \label{tail}
     \ee
as shown by the solid line. Here $q_1=2$ GeV and $\tilde q=0.4$ 
GeV.  We recall that the free $q{\overline q}$ threshold for SU(3) is
   \be
   R_{\ge 3} =\sum_i \frac{\sigma_i(\ge 3\pi)}
                          {\sigma(e^+e^-\rightarrow\mu^+\mu^-)}
   = 2 \left(\frac{N_c}{2}\sum_{q=u,d,s} e_q^2\right) =2\ ,
   \label{eq:R3}
   \ee
and underestimates the hadronic correlations by about 30\% in the 2.5-3.5
GeV region. This tail is important in the emission rates as we will note 
in the next section. It maybe interpreted as (logarithmic) corrections to
the perturbative quark result by duality. Above 3.5 GeV the charm effects
show up.

\section{Emission Rates}
\label{sec:4}

\subsection{Dielectron Rates}

\begin{figure}
\centerline{\epsfig{file=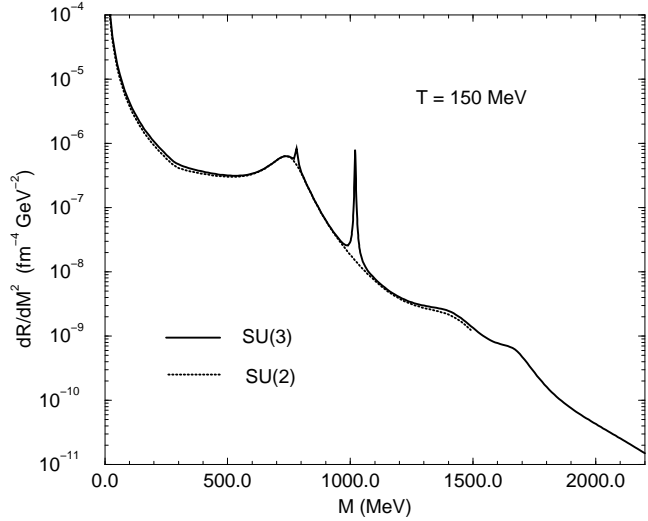,height=8cm}}
\caption{Dielectron rates at $T=150$. See text.}
\label{intrate}
\end{figure}

Given the two-point spectral weights, it is then straightforward to 
reconstruct the emission rates using Eqs.~(\ref{zero}-\ref{two}).
The contributions Eq.~(\ref{three}) are found to be numerically small
and will be ignored. ${\bf R}$ can be reexpressed in terms of
the invariant dielectron mass $M=\sqrt {q^2}$,
the rapidity $\eta$ and the magnitude of the transverse momentum
$q_t$\cite{LARRY}
     \be
     \frac{d\R}{d^2M} &=& \int dy\int d q_t^2 
       \; \frac{\pi}{2}\; \frac{d\R}{d^4 q}(M,\eta,q_t).
     \ee
In Fig.~\ref{intrate}, we show our new three flavor-rate (solid line)
in comparison to the two-flavor rate~\cite{JIM} (dashed line). Aside
from the omega and phi which were absent in the two-flavor analysis, the
results are overall consistent for $T=150$ MeV. 
The chiral reduction results show a substantial enhancement of the
low mass dielectrons in comparison to a simple PCAC treatment~\cite{JIM}.

\begin{figure}
\centerline{\epsfig{file=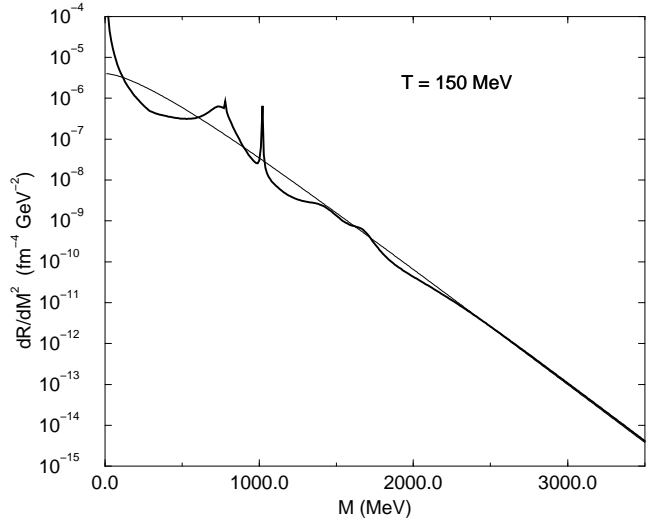,height=8cm}}
\caption{Hadronic gas (bold) versus quark gas (thin). See text.}
\label{fig:plasma}
\end{figure}

\begin{figure}
\centerline{\epsfig{file=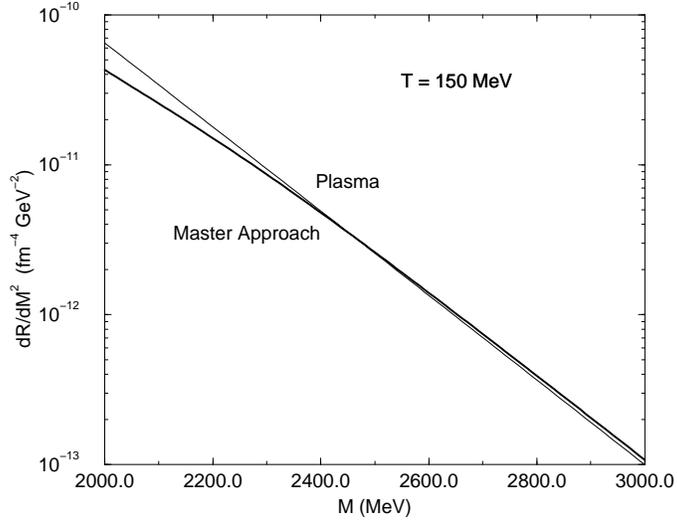,height=8cm}}
\caption{Same as Fig.~\ref{fig:plasma}.}
\label{fig:plasma2}
\end{figure}

In Fig.~\ref{fig:plasma} we show the rates from a hadronic gas 
in comparison to a free quark rate from a quark gas~\cite{HLZ}
at $T=150$ MeV. The plasma rates are large in the low mass region
(more than a factor of 2), above the phi (more than a factor of 2),
and around the 2 GeV region (about a factor 1/2).  Surprisingly,
however, the hadronic tail Eq.~(\ref{tail}) ($e^+e^-$ into hadrons)
still provide substantial emission strength even through the 
high mass region in comparison to a free quark gas at the same 
temperature. An enlargement of that region is shown in
Fig.~\ref{fig:plasma2}. 

\subsection{Photon Rates}

\begin{figure}
\centerline{\epsfig{file=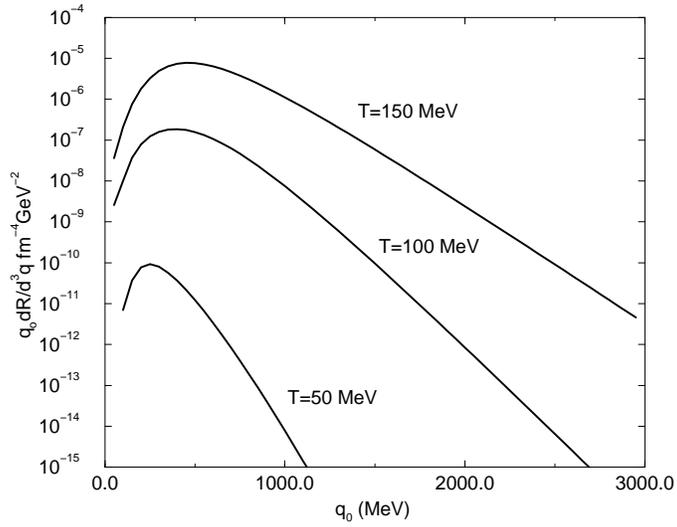,height=8cm}}
\caption{The photon emission rate from a hadronic gas.}
\label{photon_e}
\end{figure}

The rate for photons follows from Eq.~(\ref{eq:2}) at $q^2=0$,
     \be
     q^0 \frac{d\R}{d^3q} = -\frac{\alpha}{4\pi^2} \W(q).
     \ee
Fig.~\ref{photon_e} shows the photon emission rate for $T=50,100,150$ MeV.
They are overall consistent with the photon rates established for the SU(2)
case~\cite{JIM}.

\section{Emission Rates from a Fire Ball}
\label{sec:5}

In a heavy-ion collision the electromagnetic emission occurs from various
stages of the collision process. In this part, we will focus on the low and
intermediate dilepton emission rates (up to 1.5 GeV). We will assume that 
they emanate from a simple fire ball composed of a hadronic gas. The fire
ball will be modeled after transport codes~\cite{LKB96}, for the CERES
(S-Au) and HELIOS-3 (S-W) heavy-ion collisions. In particular, the expansion
will be assumed homogeneous, with a volume and temperature parametrized 
as~\cite{JIM,RAPP}
     \be
     V(t) &=& V_0\left(1+\frac{t}{t_0}\right)^3 \nn
     T(t) &=& (T_i-T_{\infty})e^{-t/\tau}+T_\infty
    \label{fire} 
    \ee
with $t_0 =10$ fm/c, $T_i = 170$ MeV, $T_\infty= 110$ MeV,
$\tau=8$ fm/c, and the value of $V_0$ is absorbed into the over-all 
normalization constant $N_0 V_0$ ($=6.76\times 10^{-7}$ fm$^3$) in 
Eq.~(\ref{eq:rate}), which is fixed by the transport results. 
The freeze-out time will be
set to $t_{f.o.}=10$ fm/c. Since we would like to only address the issue
of a thermal gas of mesons, we will ignore the baryons assuming their
contribution at present energies to be small in the leading density
approximation~\cite{JIM2}. The mesonic
fire ball is more appropriate for RHIC as opposed to CERN energies.

\begin{figure}
\centerline{\epsfig{file=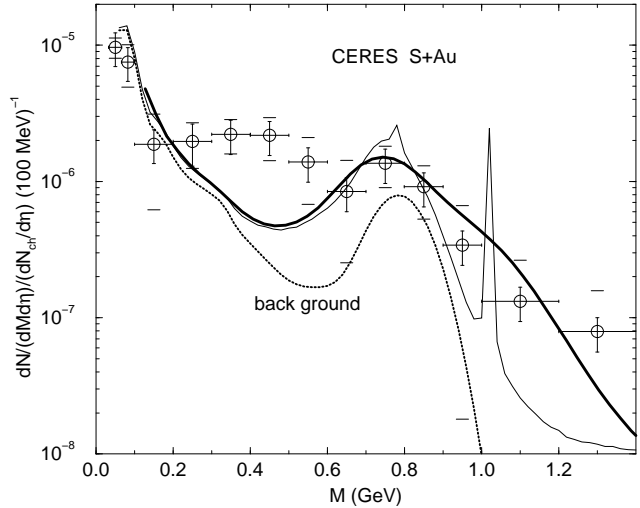,height=8cm}}
\caption{Dielectron rates for CERES S-Au experiment. See text.}
\label{fig:ceres}
\end{figure}

\begin{figure}
\centerline{\epsfig{file=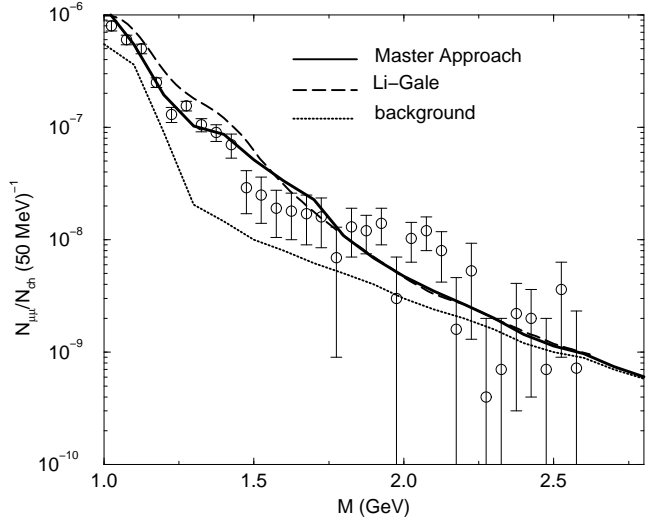,height=8cm}}
\caption{Dimuon rates for HELIOS-3 S-W experiment. See text.}
\label{fig:helios}
\end{figure}

Using Eq.~(\ref{fire}) and the above rate, the final expression for the
integrated emission rate per unit rapidity $\eta$ and invariant mass $M$
is
     \be
     \frac{dN/d\eta dM}{dN_{ch}/d\eta} (M)
      &=& N_0 M \int_0^{t_{f.o.}} dt V(t)
     \int\frac{d^3 q}{q_0} A(q_0,q^2)\frac{d\R}{d^4q} \ .
     \label{eq:rate}
     \ee
The acceptance function $A(q_0,q^2)$ enforces
the detector cut $p_t> 200$ MeV, $2.1<\eta<2.65$, and 
$\Theta_{ee}> 35$ mrad for CERES. For HELIOS-3, 
Eq.~(\ref{eq:rate}) can be used by integrating over $\eta$
with the cut $m_t\ge 4 (7-2\eta), 
m_t\ge \sqrt{(2 m_\mu)^2+(15/\cosh(\eta))^2}$. 
The results for CERES and HELIOS-3 are summarized
in Figs.~\ref{fig:ceres} and \ref{fig:helios}. 


The thin line in Fig.~\ref{fig:ceres} is our result using
only the two-point spectral weights in Eq.~(\ref{eq:rate}),
while the thick line follows from an additional gaussian 
smearing over the detector mass resolution,
   \be 
   \int_0^\infty d M^\prime\frac{1}{\sqrt{2\pi}\sigma} 
     \exp\left(-\frac{(M- M^\prime)^2}{2\sigma^2} \right) 
     \frac{dN/d\eta dM}{dN_{ch}/d\eta} (M^\prime),
   \ee
with $\sigma = 0.1 \times M$.
The background (dotted line) was taken from a transport 
model~\cite{LKB96,CHARLES}. The $\phi$ resonance is clearly visible.
The data are from~\cite{CERES}. Clearly the fire-ball evolution
together with the mesonic gas do not account for the low mass
dileptons. In Ref.~\cite{JIM}, it was shown that only large nucleonic
densities could account for the data to leading order in the 
hadronic densities. 

The solid line in Fig.~\ref{fig:helios} is our three-flavor
result for the HELIOS-3 experiment. The dashed curve refers to
the results discussed recently by Gale and Li~\cite{CHARLES}
using effective Lagrangians and a variety of two-body reactions.
Our results are in overall agreement with theirs. The dotted line
is again the background contribution from the transport 
model~\cite{LKB96,CHARLES}. The fair agreement of the fire ball 
with the dimuon spectra above 1 GeV indicates that a thermal hadronic gas 
treatment is consistent with the data. Some enhancement in the low 
mass region maybe achieved by adding baryons, but not enough in
our leading density approximation to bring it into agreement with
the data~\cite{JIM}.

\section{CERES with $p_t$ cut}
\label{sec:6}

\begin{figure}
\centerline{\epsfig{file=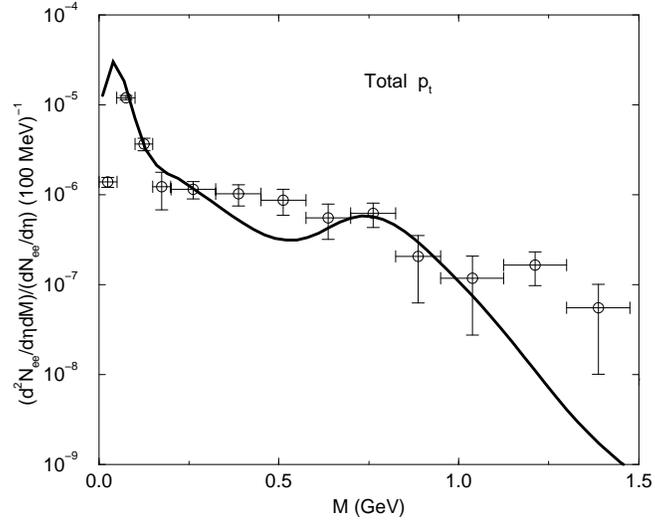,height=8cm}}
\caption{Dielectron rates for the total $p_t$ of
CERES Pb-Au experiment.}
\label{fig:total_pt}
\end{figure}

\begin{figure}
\centerline{\epsfig{file=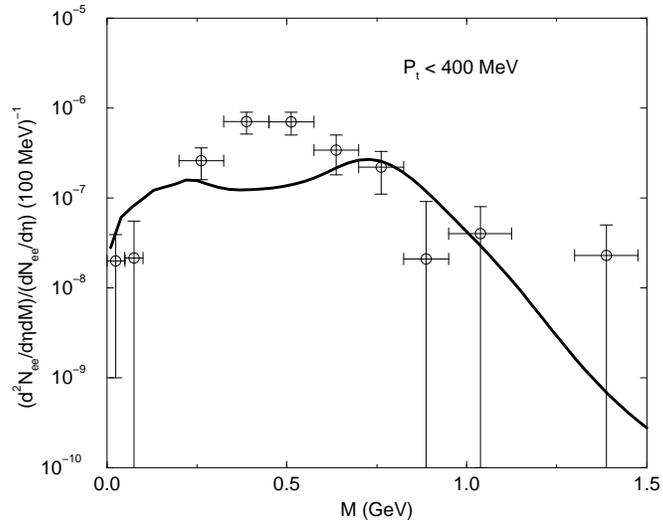,height=8cm}}
\caption{Dielectron rates for low $p_t$ of CERES Pb-Au experiment.}
\label{fig:low_pt}
\end{figure}

Recently, the CERES collaboration analyzed the $p_t$ dependence
of the dielectron pairs~\cite{WURM}. Using our mesonic rates and
the fire ball evolution Eqs.~(\ref{fire}-\ref{eq:rate}) we show in 
Figs.~\ref{fig:total_pt} and \ref{fig:low_pt} our results in 
comparison to data. The background contribution in our case was 
borrowed from transport calculations~\cite{LI_pt}. 
The initial temperature of the fire ball is set at $T_i=160$ MeV
and the freeze out temperature at $T_\infty =105$ MeV for which
the time scale is $t_{f.o.}= 20$ fm/c, with $\tau=10$ fm/c and 
$t_0=10.8$ fm/c. The new normalization constant, $N_0 V_0 = 
3.45\times 10^{-7}$ fm$^3$, is fixed by the transport 
results~\cite{LI_pt}.

\begin{figure}
\centerline{\epsfig{file=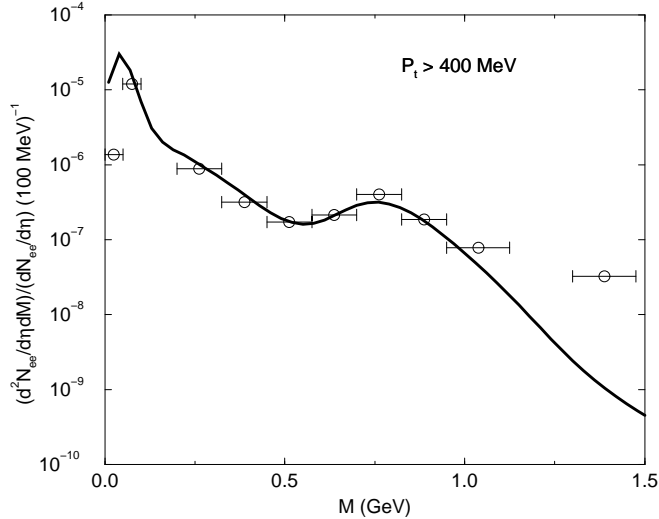,height=8cm}}
\caption{Dielectron rates for high $p_t$ of CERES Pb-Au experiment.}
\label{fig:high_pt}
\end{figure}

Our analysis based on a baryon free hadronic gas (with strange
mesons) does not reproduce the low mass dielectron enhancement
in both Fig.~\ref{fig:total_pt} and \ref{fig:low_pt}. 
Since most of the discrepancy of the low mass enhancement comes from low
$p_t$ contribution, taking into account the large statistical and
systematical errors for $M>1.0$ GeV, we find that our high $p_t$ spectrum is 
consistent with the data. In Fig.~\ref{fig:high_pt}, the data are the
differences between the mean values of the total and low $p_t$ spectrum,
so we do not include the vertical error bars.

\section{Conclusions}
\label{sec:7}

We have extended the dilepton and photon emission rates from
a thermalized gas of hadrons to account for strange mesons without
baryons. The analysis relies on the extension of the chiral reduction
formulas from SU(2) to SU(3). The hadronic gas now includes the
effects from kaons, etas and phis. As expected, only substantial 
changes in the rates are seen around the omega and phi region.

In the low mass region there is a substantial enhancement due to
the mixing with the axial particles, but not enough to account
for the CERES data after time-evolution and detector cuts. In the
intermediate mass region, there is good agreement with the HELIOS 
data. The baryonic effects are important in the low mass region,
but not enough to account for the discrepancy seen in the Pb-Au
collisions at CERN within the density expansion~\cite{JIM}. The
large and persistent hadronic tails make the thermal mesonic emission
rates comparable to the ones from a thermal gas through the intermediate
mass region.

As we have indicated in the introduction, our analysis is not
based on a model. Our only assumption is that the emission rates
follow from a baryon free and dilute hadronic gas. The rest follows
from symmetry and data. As such all model calculations should be in
agreement with our results under these assumptions, and it is gratifying
to see that the recent analysis performed by Gale and Li~\cite{CHARLES} using
reaction rates does.

The fact that the HELIOS-3 data can be explained without recourse to 
additional assumptions, make part of our arguments reliable. The
persistent disparity in the low mass region with CERES and the newly
measured $p_t$ spectra may indicate the need for an enlargement of the 
original assumptions. The importance of the thermal quark gas emissivities in
the 1/2 and 1 GeV region maybe indicative of some simple nonperturbative
effects in the partonic phase for the temperatures considered~\cite{HZ90}. 
This issue will be addressed next \cite{HLZ}.

\section*{Acknowledgements}

We have benefitted from several discussions with Gerry Brown, Axel Drees,
Charles Gale, Guoqiang Li, Madappa Prakash, Ralf Rapp, Edward Shuryak,
Heinz Sorge and  Jim Steele. 
This work was supported in part by the U.S. Department of Energy under 
Grant No. DE-FG02-88ER40388.

\end{document}